\documentclass[showpacs,aps,prl,twocolumn,superscriptaddress]{revtex4-1}

\pdfoutput=1

\usepackage{amsmath,amsfonts,amssymb,bm,graphicx,hyperref,color}

\begin{document}

\title{Comment on ``Spontaneous liquid-liquid phase separation of water'' by T. Yagasaki, M. Matsumoto and H. Tanaka, Phys. Rev. E \textbf{89}, 020301 (2014) }

\author{David T. Limmer} 
\email[Corresponding author, ]{dlimmer@princeton.edu}
\affiliation{Princeton Center for Theoretical Science, Princeton University, Princeton NJ 08544 USA}
\author{David Chandler}
\email[Corresponding author, ]{chandler@berkeley.edu}
\affiliation{Department of Chemistry, University of California, Berkeley, CA 94720, USA}


\begin{abstract}
Yagasaki et al. present results from a molecular dynamics trajectory illustrating coarsening of ice, which they interpret as evidence of transient coexistence between two distinct supercooled phases of liquid water.  We point out  that neither two distinct liquids nor criticality are demonstrated in this simulation study.  Instead, the illustrated trajectory is consistent with coarsening behaviors analyzed and predicted in earlier works by others. 
\end{abstract}

\maketitle

With figures and an illuminating video, Yagasaki et al.~ \cite{Yagasaki2014} present interesting and instructive results for a 1 $\mu$s molecular dynamics trajectory of ice coarsening from its supercooled liquid.  They describe the results as if to provide evidence for the existence of two distinct liquid phases of water, one of high density and the other of low density, but more straightforwardly, the results illustrate coarsening behaviors that we and others have analyzed in earlier work~\cite{limmer2013a, limmer2013b, molinero2013}.  This Comment is written to counter the suggestion that Ref.~\cite{Yagasaki2014} provides evidence contrary to that earlier work.

The phenomena in question are established in Ref.~\cite{Yagasaki2014} by simulating water with a closed, rigid, periodically replicated box of 4000 classical particles interacting through a variant of the ST2 potential energy function.  The phase behaviors of this and related models have been examined by many.  Our recent paper~\cite{limmer2013a} as well as Ref.~\cite{Yagasaki2014} provide references.  The ST2 model, like any reasonable model of water, exhibits first-order phase transitions between liquid and ice, and for a range of temperatures below the melting temperature, $T_\mathrm{m}$, the liquid can exist as a metastable phase.  Below a yet lower temperature, $T_\mathrm{s}$, the liquid is unstable with respect to crystallization.  Preparation of the liquid at a temperature $T \lesssim T_\mathrm{s}$ leads directly to coarsening, which takes place in stages, the last of which occurs for the ST2 model on scales of microseconds~\cite{same2013}. 

The illustrative 1$\mu$s trajectory~\cite{Yagasaki2014} is initiated with a typical liquid configuration from $T > T_\mathrm{m}$ and instantly quenched to $T \approx T_\mathrm{s}$.  Because volume and number of molecules are held fixed in the simulation, the subsequent coarsening ultimately reaches a state with two domains separated by a sharp stationary interface.  One domain is ordered ice, the other is coexisting liquid.  The density of the former is lower than that of the latter. Before reaching that final state, the system exhibits large fluctuations, as is expected of a system prepared far from equilibrium.  

For most of the trajectory, the box is filled by two liquid domains, with interfacial widths and shapes changing markedly with time.  The domains appear at the first stages of the trajectory.  The density of one is about 10\% lower than that of the other.  The ordered crystal eventually emerges from the lower-density domain, consistent with the findings of Ref.~\cite{molinero2013}.  Long-ranged order appears only late in the trajectory because such ordering occurs on times scales about 100 times longer than those for which most density fluctuations occur -- the specific extent of time-scale separation being model dependent~\cite{limmer2013a, limmer2013b}.  For ST2 in particular, prior analysis of the early stages of coarsening already predicts the time scales of fluctuations now illustrated by Yagasaki et al.'s trajectory.  See, e.g., Fig. 1 of Ref.~\cite{Yagasaki2014} and Fig. 4 of Ref.~\cite{limmer2013a}.  In both the new simulations and the prior theoretical analysis, liquid domains convert from one density to the other on a time scale of tens of nanoseconds.  In both the new simulations and the prior theoretical analysis~\cite{limmer2013b}, long-range order (i.e., late stage coarsening) develops after 1 $\mu$s.  

Thus, there is general agreement on the time scales and amplitudes of density fluctuations and ultimate ordering of supercooled ST2 water.   What distinguishes Ref.~\cite{Yagasaki2014} is its vivid quality, which allows us to inspect the notion that this model of supercooled water, and perhaps water itself~\cite{stanley}, can exist as two coexisting liquids with an associated critical point.   
To qualify as distinct phases, and not simply transient non-equilibrium fluctuations, interfaces separating domains must have a finite surface tension.  But attempts to compute a liquid-liquid surface tension~\cite{limmer2013a} find instead only one liquid phase of water \cite{nature2014}.
Indeed, a finite surface tension would inhibit the large interfacial fluctuations observed in Ref.~\cite{Yagasaki2014}.  Zero surface tension is possible, but only at a critical point.  The transient structures observed in Ref.~\cite{Yagasaki2014} contain only two large domains, not a collection of many domains of various sizes as would be required of criticality.  Rather than coexistence or criticality, therefore, the results of Yagasaki et al.~\cite{Yagasaki2014} are entirely consistent with those of English et al.~\cite{english2013} who suggest that stable interfaces do not exist between putative high- and low-density phases of supercooled water.  

Consistency with time scales predicted for coarsening~\cite{limmer2013a, limmer2013b} lends support to viewing the behaviors shown in Ref.~\cite{Yagasaki2014} in terms of dynamics.  Further, as noted by Binder~\cite{Binder2014}, finite lifetimes of metastable states make it impossible for fluctuations of early-stage coarsening to be like criticality between two distinct liquid phases in pseudo equilibrium.
Viewing ice coarsening on its own terms, as a far-from-equilibrium behavior, is straightforward by contrast, and with the rendering of Ref.~\cite{Yagasaki2014} we can see how ice coarsening occurs better than ever before.

\vspace{2mm}

\end{document}